\documentclass[useAMS,usenatbib,usegraphicx]{mn2e}
\usepackage{times}
\usepackage{amssymb}
\def \HI {H {\small I} }

\title[Velocity statistics from spectral line data]{Velocity statistics from spectral line data: effects of density-velocity correlations, magnetic field, and shear}
\author[Esquivel et al.]{A. Esquivel,$^1$\thanks{E-mail: esquivel@astro.wisc.edu; lazarian@astro.wisc.edu; pogosyan@phys.ualberta.ca;
cho@astro.wisc.edu}, A. Lazarian,$^1$\footnotemark[1] D. Pogosyan,$^2$\footnotemark[1]
J. Cho$^1$\footnotemark[1] \\
$^1$Astronomy Department, University of Wisconsin-Madison, 475 N. Charter St., Madison, WI 53706, USA \\
$^2$Department of Physics,University of Alberta, Edmonton, Alberta T6G 2J1, Canada}

\begin{document}

\date{Draft Version \today}

\maketitle

\begin{abstract}
In a previous work Lazarian and Pogosyan suggested a technique to extract velocity and density statistics, of interstellar turbulence, by means of analysing statistics of spectral line data cubes. In this paper we test that technique, by studying the effect of correlation between velocity and density fields, providing a systematic analysis of the uncertainties arising from the numerics, and exploring the effect of a linear shear. We make use of both compressible MHD simulations and synthetic data to emulate spectroscopic observations and test the technique. With the same synthetic spectroscopic data, we also studied anisotropies of the two point statistics and related those anisotropies with the magnetic field direction. This presents a new technique for magnetic field studies. The results show that the velocity and density spectral indices measured are consistent with the analytical predictions. We identified the dominant source of error with the limited number of data points along a given line of sight. We decrease this type of noise by increasing the number of points and by introducing Gaussian smoothing. We argue that in real observations the number of emitting elements is essentially infinite and that source of noise vanishes.
\end{abstract}


\section{Introduction}

The very large Reynolds numbers (defined as the ratio of the inertial force to the viscous force acting on a parcel of gas) clearly suggest that the interstellar medium is turbulent. Understanding of this turbulence is crucial to the correct and complete description of many physical processes that take place in the ISM: molecular cloud dynamics, star formation, heat transfer, magnetic reconnection, accretion disks, cosmic ray propagation and diffusion, just to mention a few.

To adequately describe turbulence we need to use statistical methods, to extract the underlying regularities and reject incidental details. So far, studies of turbulence statistics have been most successful using interstellar scintillations, and have provided important information of the density statistics on scales $10^8-10^5 cm$, (see \citealt{NG, SG}). These studies are done with the ionised media, and are restricted to the density fluctuations. We expect to observe power-law power spectrum of density in a turbulent medium, however density fluctuations alone fail to distinguish between fossil and active turbulence. The distribution of sizes of sand grains in a beach follows a power-law but it is not turbulence. Therefore would be better to study a more direct and dynamically important statistic: velocity fluctuations. Studies of the neutral media have included velocity information, e.g. using spectral line widths, centroids of velocity \citep{MB, S87}; or more fancy statistics \citep{HS, R99, BHa, BHb}.

Turbulence statistics is a very important link between theory and observations. Recent breakthroughs in turbulence studies have been possible with the use of scaling laws (see reviews by \citealt*{CL03,CLV03}, and references therein). Scaling laws predictions can be confronted with observations.

A discussion of various approaches to the study of turbulence with spectral line data can be found in \citet{L99}. In particular the problem of the contribution of velocity and density fluctuations to the emissivity statistics was addressed in \citet[][hereafter LP00]{LP00}, where analytical description of the power spectrum of the emissivity in velocity channels was obtained. This study provides a technique to extract statistics of density and velocity from spectral line data cubes. \HI has been chosen as a test case because it was possible to disregard self-absorption, and consider the emissivity linearly depending on the density (see review in \citealt*{LPE02}). The technique, termed as velocity-channel analysis (or simply, VCA), is based on the variations of the power spectra in velocity channels with the change of velocity resolution. Predictions in  LP00 have been confirmed through observations \citep{SL}. The first tests using numerical data in \citet{LPVP} also confirmed the predictions in LP00. However they did not address the problems of numerical noise, effects of magnetic field and galactic shear. Moreover the study of velocity-density correlations was performed over a very limited dynamical range. 

In this work we subject the predictions in LP00 to further scrutiny. In particular, we systematically analyse the effects of velocity-density correlations, and unlike \citet{LPVP} we include a study of the effect of shear. In addition we address the issue of an anisotropic turbulent cascade, with their implications on the VCA. We also analyse the sources of numerical noise.

In \S $2$ we review the basic problem, we then (\S $3$) test the extraction of the spectral indices using the VCA, and obtain the density-velocity correlations. In \S $4$ we study the effect of shear in the VCA. The anisotropic cascade, and a possible new method to extract information of the magnetic field direction can be found in \S $5$. And we finally provide a summary of our results (\S $6$). 

\section{Statistics of spectral line data and VCA}

Spectral line observations contain important information of density and velocity. The problem is that both velocity fluctuations and density fluctuations contribute to the intensity of a spectral line at a given velocity, and their separation is far from trivial.

\subsection{3D turbulence statistics}

Turbulence, due to its stochastic nature is studied using statistical tools, such as 3D correlation and structure functions (see \citealt{MY}). For instance, the density correlation function can be expressed as
\begin{equation}
\label{cf}
\xi(\bmath{r})=\langle \rho(\bmath{x})\rho(\bmath{x}+\bmath{r})\rangle,
\end{equation}
where $\bmath{x}$ is the spatial position ($xyz$ in Cartesian coordinates), $\bmath{r}$ is a spatial separation (or `lag'), and the angular brackets denote an average to be performed over all $\bmath{x}$ space. The power spectrum provides an alternative description and is related to the correlation function (CF) as
\begin{equation}
\label{def_ps}
P(\bmath{k})=\int \rmn{d}\bmath{r}\, \rmn{e}^{i\, \bmath{k}\bmath{\cdot}\bmath{r}}\xi(\bmath{r}),
\end{equation}
where $\bmath{k}$ is the wave-number, related to the scale under study  like $k=2\pi /r$, and the integration is performed over all 3D space. For power-law statistics, the $N$ dimensional power spectrum ($P_N\propto k^{n}$) and the correlation ($\xi_N \propto r^{m}$) have indices related as $n=-N-m$, in other words\footnote{Notice that the spectral indices contain the sign. This notation is unfortunate (contrary to a conventional spectral index of the form $P_N \propto k^{-n}$, with $n$ positive). However, it was introduced in LP00, and for consistency we adhere to it.}
\begin{equation}
\label{indices_rel}
(spectral~index)=(-spatial~dimensions-CF~index).
\end{equation}
To describe velocity fields is customary to use structure functions:
\begin{equation}
\label{eq:sf}
SF(\bmath{r})=\langle [\bmath{v}(\bmath{x})-\bmath{v}(\bmath{x}+\bmath{r})]^2 \rangle.
\end{equation}
They provide a more formal treatment of velocity fields (\citealt{MY}). In the case of isotropic velocity fields it differs from the correlation function only by a constant. Therefore for power-law statistics it has the same spectral index $m$ as the correlation function (i. e. $SF \propto r^{m}$). For instance, Kolmgorov turbulence scales as $v \propto r^{1/3}$, which corresponds to $m=2/3$.

\subsection{Statistics in position-position-velocity (PPV) space}

In spectroscopic observations we don't observe the gas distribution in real space coordinates $\bmath{x}\equiv (x,y,z)$. Instead the intensity of the emission in a given spectral line is measured towards some direction in the sky, given by $\bmath{X}\equiv (x,y)$, and at a given line-of-sight (LOS) velocity $v$. Thus, the coordinates of observational data are $(\bmath{X},v)$. Therefore, observational data are contained in, so called, position-position-velocity (or simply `PPV') cubes. If we identify the $z$ coordinate with the LOS, then the relation between the real space and PPV descriptions is that of a map $(\bmath{X},z)\rightarrow(\bmath{X},v)$.

At a given position $\bmath{x}$ the LOS component of the velocity $v$ can be decomposed as a regular flow velocity $u_{reg}(\bmath{x})$,  plus a thermal component $v_{thermal}$ and a turbulent component $u(\bmath{x})$. This way the observed Doppler shifted atoms follow a Maxwellian distribution of the form
\begin{equation}
\label{max}
\phi_v(\bmath{x})\rmn{d}v=\frac{1}{(2\pi\beta)^{1/2}}\exp \left\{-\frac{[v-v_{reg}(\bmath{x})-u(\bmath{x})]^2}{2\beta}\right\} \rmn{d}v,
\end{equation}
where $\beta=\kappa_BT/m$, $m$ is the atomic mass, $T$ the gas temperature, and $\kappa_B$ the Maxwell-Boltzmann constant.

The density of emitters in PPV space $\rho_s(\bmath{X},v)$ can be obtained by integration along the LOS:
\begin{equation}
\label{emissivity}
\rho_s(\bmath{X},v)\rmn{d}\bmath{X}\rmn{d}v=\left[ \int \rmn{d}z~\rho(\bmath{x})\phi(\bmath{x})\right] \rmn{d}\bmath{X}\rmn{d}v,
\end{equation}
where $\rho(\bmath{x})$ is the mass density of the gas in spatial coordinates. This expression just counts the number of atoms along the LOS, at a given position $\bmath{X}$ with a $z$ component of velocity in the interval $[v,v+\rmn{d}v]$, and the limits of integration are defined by the extent of spatial distribution of emitting gas.

The statistics available in PPV are the correlation functions, $\xi_s\equiv \langle \rho_s(\bmath{X_1},v_1)\rho_s(\bmath{X_2},v_2)\rangle$, and the power spectra (this can be 1D, 2D or 3D) of emissivity fluctuations. Relations between this statistics and the underlying velocity and density spectra were established in LP00, where the 2D power spectrum in velocity channels was used. This spectrum can be directly obtained with radio-interferometric observations (see \citealt{L95}).

\subsection{Velocity-channel analysis (VCA)}

This technique was developed in LP00 to extract the velocity and density spectral indices of turbulence from spectral line data. LP00 provide analytical predictions for the emissivity power spectral indices in velocity channels, as a function of the velocity resolution employed. The relative contribution to the intensity in a velocity channel (also referred as `slice') to the total intensity fluctuations changes in a regular fashion with the width of the velocity slice. This is easy to understand qualitatively since more velocity fluctuations within a channel are averaged out as we increase the thickness of the channel, decreasing the relative contribution of velocity. 

The notation that we will use is such that: $n$ is the 3D density spectral index ($P_{\rho}\propto k^n$); $\mu$ the 3D velocity spectral index ($P_{v}\propto k^{\mu}$); $\gamma$ corresponds to the spectral index in velocity slices ($P_{n}\propto k^{\gamma}$); and $m$ is the structure function index of the velocity. From eq. (\ref{indices_rel}) is clear that 
\begin{equation}
\label{mu}
\mu=-3-m.
\end{equation}

For a given 3D density power spectrum two distinct regimes are present when: a) $n>-3$ (`shallow' density case), and (b) $n<-3$ (`steep' density case). A summary of the predictions in LP00 for the intensity spectral index in 2D velocity channels is given in Table \ref{tab:LP00}.
\begin{table*}
\centering
\begin{minipage}{120mm}
\caption{A summary of analytical results derived in LP00.}
\label{tab:LP00}
\begin{tabular}{lcc}
\hline
Slice & Shallow 3-D density & Steep 3-D density\\
thickness & $P_{n} \propto k^{n}$, $n>-3$&$P_{n} \propto k^{n}$, $n<-3$\\
\hline
2-D intensity spectrum for thin\footnote{channel width $<$ velocity dispersion at the scale under study}~slice & $\propto k^{n+m/2}$ & $\propto k^{-3+m/2}$\\
2-D intensity spectrum for thick\footnote{channel width $>$ velocity dispersion at the scale under study}~slice & $\propto k^{n}$ & $\propto k^{-3-m/2}$  \\
2-D intensity spectrum for very thick\footnote{substantial part of the velocity profile is integrated over}~slice & $\propto k^{n}$ & $\propto k^{n}$\\
\hline
\end{tabular}
\end{minipage}
\end{table*}
In both cases (shallow or steep) the power-law index {\it `gradually steepens'} with the increase of the velocity slice thickness. In the thickest velocity channels all the velocity information is averaged out and we naturally get the density spectral index $n$. On the other hand, the velocity fluctuations dominate in thin channels. For instance, in the case of a steep density ($n<-3$) the structure seen in the velocity channels is determined only by velocity fluctuations (is not explicitly dependant on the index $n$). This immediately sends a warning against identifying structures in PPV as actual density enhancements (`clouds'). The spectral velocity index can be obtained if we combine the density velocity index from the very thick channels and the structure function index $m$ from thin channels (using Table \ref{tab:LP00} and eq. \ref{mu}). Notice that the notion of thin and thick slices depends on the turbulence scale under study, and the same slice can be thick for small-scale turbulent fluctuations and thin for large scale ones. The formal criterion for the slice to be thick is that {\it the dispersion of turbulent velocities on the scale studied should be less than the velocity slice thickness}. Otherwise the slice is {\it thin}.

From Table \ref{tab:LP00} one may observe that for any given density spectral index, the spectrum of fluctuations becomes shallower as the velocity field gets steeper. This is somewhat counterintuitive. But can be understood as follows: a steeper velocity index corresponds to relative more power on the largest scales.
A steeper velocity power-spectrum means a larger velocity difference across scales, and correspondingly to a larger velocity dispersion. In PPV space, large eddies, because their velocity dispersion, will be spread over a larger velocity range than smaller eddies. In individual channels, the number of emitters associated with large eddies will decrease relative to that of small eddies as the velocity contrast increases. Therefore, in velocity slices, the power at small $k$ numbers (large scales, from large eddies) decreases for a steeper velocity, rendering in a shallower power-spectrum.

In following sections we perform tests of the VCA using numerical data, and take a step further exploring the sources of uncertainty of and the effects of shear in the VCA.

\section{Testing the VCA}

Numerical simulations provide good means of theory testing. Using numerical data, we can measure directly the velocity and density statistics, and we can produce synthetic spectra to analyse them as an observer would do. However, we need to be aware of the differences of synthetic and real data. For instance, only a limited number of points is available with numerics. This, as we show later, results in the most important source of uncertainty.

\subsection{The data}

Along the paper we make use of three types of data to construct the PPV cubes and analyse the power spectrum:

\begin{enumerate}
\item MHD simulations.
We use a 3rd-order hybrid essentially-non-oscillatory (ENO) code to simulate fully developed turbulence. It was performed on a $216^{3}$ Cartesian grid, it is isothermal, compressible, and with a mean magnetic field in the $x$ direction. To reduce spurious oscillations near shocks, we combine two ENO schemes. When variables are sufficiently smooth, we use the 3rd-order weighted ENO scheme \citep{JW99} without characteristic mode decomposition. When opposite is true, we use the 3rd-order Convex ENO scheme \citep{LO98}. We use a three-stage Runge-Kutta method for time integration. We solve the ideal MHD equations in a periodic box:
\begin{eqnarray}
\frac{\partial \rho}{\partial t} + \nabla \bmath{\cdot} (\rho \bmath{v}) =0, \nonumber \\
\frac{\partial \bmath{v}}{\partial t} + \bmath{v}\bmath{\cdot} \nabla \bmath{v} + \rho^{-1} \nabla(c_s^2\rho) - (\nabla \bmath{\times} \bmath{B})\bmath{\times} \bmath{B}/4\pi \rho =\bmath{f}, \\
\frac{\partial \bmath{B}}{\partial t} - \nabla \bmath{\times} (\bmath{v} \bmath{\times}\bmath{B}) =0,\nonumber
\end{eqnarray}
with $ \nabla \bmath{\cdot} \bmath{B}= 0$ and an isothermal equation of state. Here $\bmath{f}$ is a random large-scale driving force, $\rho$ the density, $c_s$ the sound speed, $\bmath{v}$ the velocity, and $\bmath{B}$ the magnetic field. The rms velocity $\delta V$ is maintained to be approximately unity, so that $\bmath{v}$ can be viewed as the velocity measured in units of the rms velocity of the system, and $\bmath{B}/\sqrt{4 \pi \rho}$ as the Alfv\'{e}n speed in the same units. The time $t$ is roughly 
in units of the large eddy turnover time ($\sim L/\delta V$) and the length in units of $L$, the scale of the energy injection. The magnetic field consists of the uniform background field and a fluctuating field: $\bmath{B}= \bmath{B}_0 + \bmath{b}$. The Alfv\'{e}n velocity of the mean field is roughly the same as the rms velocity. The average Mach number is $\sim 2.5$.

The outcomes are the density and velocity fields. The PPV cubes constructed with this data have a correlation between the velocity and the density fields which is self-consistent with MHD evolution. These and are used to test the applicability of the VCA where the basic assumption was to disregard such correlation.
\item `Modified' data. Due to numerical limitations, the power spectrum of the simulations mentioned above doesn't show exact power-law behaviour over the desired inertial range, making difficult to test the VCA, which deals with power-law spectra. For that reason we apply the same procedure as in \citet{LPVP} to modify the power spectrum to follow a power-law preserving most of the phase correlations. The procedure consists in replacing the amplitudes of the Fourier transform of the data so they follow a power-law while keeping the phases intact (were most of the spatial information is). The set of simulations used here show a larger well-developed inertial range than those in \citet{LPVP}; therefore the `correction' introduced is smaller, especially for the $-11/3$.
\item Gaussian fields. We use a standard procedure to generate Gaussian fields with prescribed power-law spectrum (see \citealt*{BKP}). We generate a Fourier representation of the field\begin{equation}
\label{gaussian}
F(k)=\sum k\, a_{k}\, P(k)^{1/2}\, {\rm e} ^{ikr},
\end{equation}
 where $a_{k}$ are independent Gaussian variables with dispersion $1$, and a mean of $0$ for velocity fields or $1$ for density fields; in our notation the wave-number $k=2\pi L/\lambda$, where $\lambda$ is the wavelength associated with $k$, and $L$ is the box size. To obtain the field in configuration space a Fourier transform is applied. Velocity and density fields generated this way are entirely uncorrelated. However the facility of generating as many cubes as we want in a short amount of time is useful when we want to improve the statistics of our sample.
\end{enumerate}

Comparing (ii) and (iii) allows to test the importance of density-velocity correlations.

With the data cubes mentioned above we produced synthetic spectra (i. e. PPV `cubes'\footnote{Strictly speaking they are parallelepipeds as the sizes need not be the same in all directions.}) to analyse them using VCA. The procedure to construct the PPV cubes is as follows: We generate a matrix of the same dimensions in $\bmath{X}$ as the $\bmath{x}$ data and a given number of velocity bins (velocity channels) as the third dimension. The velocity bins are equally spaced ranging from the minimum value of the velocity ($v_{min}$) to the maximum value of the velocity ($v_{max}$) in the whole velocity data cube. We then sum all the elements of the density cube in a position $\bmath{X}$ that correspond to each velocity bin. This way we simulate observational data, neglecting self absorption, and assuming that the thermal width of the emitting gas is always smaller than the thickness of the velocity channels. An introduction of an intrinsic width (equivalent to thermal) to the emitters is discussed later.

\subsection{Density and velocity correlations}

\begin{figure}
\includegraphics[width=84mm]{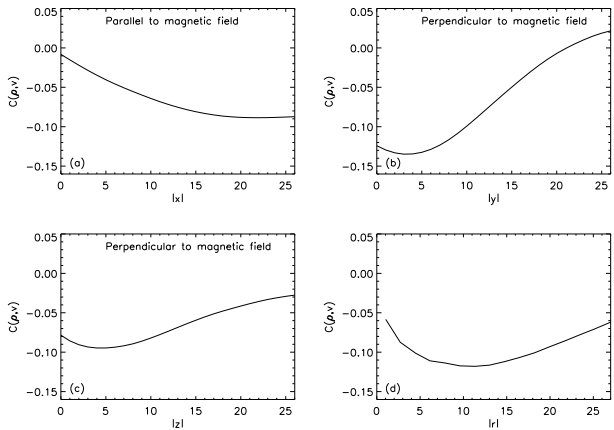}
\caption{Correlation of the density and velocity fields. (\emph{a}) the average projection of the cross-correlation on the $x\protect$ axis (parallel to the mean magnetic field), (\emph{b}) the average projection of the correlation on the $y$ axis (perpendicular to the mean magnetic field), (\emph{c}) the average projection of the correlation on the $z$ axis (perpendicular to the mean magnetic field, and (\emph{d}) average correlation as function of the spatial separation $r$.
\label{fig:rho_v_corrs}}
\end{figure}

LP00 considered density and velocity either uncorrelated or correlated to maximal degree. What would happen to intermediate cases? We computed the correlation of the density and the velocity fields as
\begin{equation}
\label{corr}
C(\bmath{r})=\frac{\langle \rho (\bmath{x})\bmath{u}(\bmath{x}+\bmath{r})\bmath{\cdot} \bmath{r}/|\bmath{r}|\rangle }{\sigma _{\rho }\sigma _{\bmath{u}}}.
\end{equation}
Where $\bmath{r}$ is the spatial separation (`lag'), $\bmath{x}$ is the spatial position, $\rho$ the density field, $\bmath{u}$ is the velocity field, $\sigma _{\rho }$ and $\sigma _{\bmath{u}}$ are the standard deviations of the density and velocity fields respectively. The average is to be performed over all $\bmath{x}$ space and angles of $\bmath{r}$. For numerical economy \citet{LPVP} used Fourier transform techniques to obtain similar information (although restricted to the LOS velocity component). Here this is done directly using equation (\ref{corr}), which is applied to the original output from the MHD simulations. We found moderate correlations, the maximum correlation (in absolute value) was of $\sim 0.22$. This correlation is shown in Fig. \ref{fig:rho_v_corrs}, where we plotted the average projection of the cross-correlation on the $x$, $y$, $z$ axis and the average correlation as a function of the separation $r$.
Clearly the velocity and density correlations are very different along and perpendicular to the magnetic field direction. Some implications of this anisotropy are discussed in \S 5.

\subsection{Applying the VCA}

In this subsection we will start the test of the VCA generating several data cubes with known (`modified') spectral indices, and arbitrary velocity resolutions that will be shown to range from very thick, to excessively thin. This is to check the general trends predicted in LP00. Then with a more careful choice of the velocity thickness of the channel, we will narrow down the measured spectral indices to check the numerical value with the analytical expectation. A more detailed analysis of the sources of error introduced with numerics, with an example case of Kolmogorov indices, follows in the next subsection.

\begin{figure*}
\includegraphics{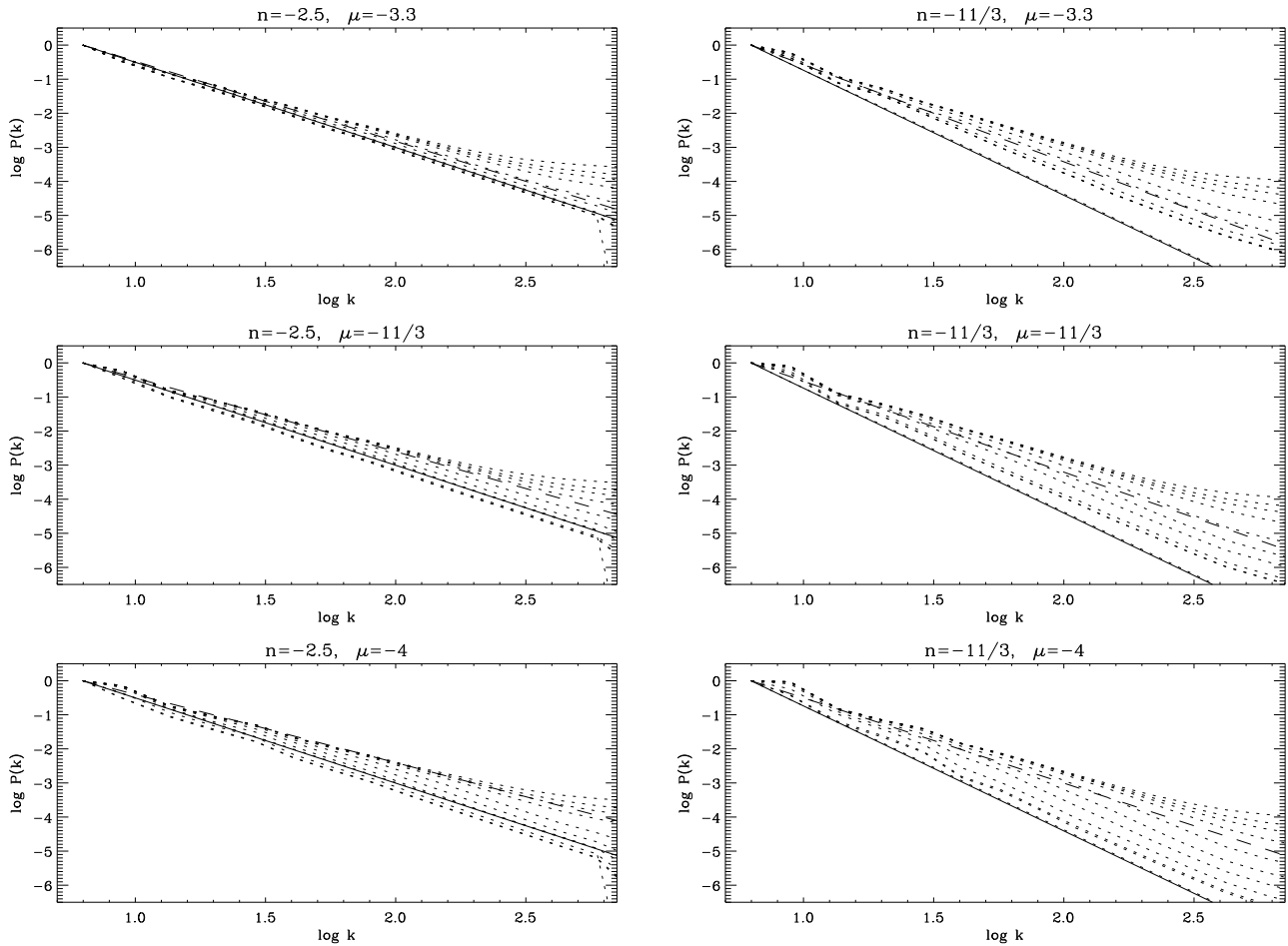}
\caption{Power spectra in velocity channels for simulated observations with density spectral indices of $n=-2.5$ and $n=-11/3$, with  velocity spectral indices of $\mu =-3.3$, $11/3$, and $-4$ (as labelled in the title of each plot). In each panel the \emph{dotted} lines correspond to the power-spectra in velocity channels for PPV cubes generated with,  from top to bottom $150$, $100$, $75$, $50$, $25$, $15$, $10$, $6$, $4$, $2$, and $1$ channels respectively. For reference we plotted the predictions in LP00: in \emph{dashed} lines are the predictions for thin channels, in \emph{solid} lines for very thick channels.}
\label{fig:ps_rho_11_3_vz_all}
\end{figure*}

We produced density data cubes with spectral indices modified to $-2.5$ and $-11/3$, LOS velocity cubes with indices of $-3.3$, $-11/3$ (Kolmogorov), and $-4.0$. With each velocity and the density fields we constructed PPV cubes with velocity resolutions, ranging from $150$ to $1$ number of channels. The 2D power spectra in velocity channels (averaged over all velocity channels for each PPV cube) is presented in Fig. \ref{fig:ps_rho_11_3_vz_all}.

We should stress, that whether a slice in velocity space can be considered
{\it thin} or {\it thick} depends not only on the slice width $\delta v=(v_{max}-v_{min})/N=\Delta v/N$ (where $N$ is the number of channels) but on the scale as well. For power-law statistics the squared dispersion, $\sigma_L^2$, over the box size, $L$, equals $CL^m$ (here $m$ is the velocity structure function index related to the spectral index as described in eq \ref{indices_rel}); similarly the velocity dispersion squared, $\sigma_r^2$, over the scale $r$ is $Cr^m$. Therefore the criteria for a channel
to be considered thin ($\delta v < \sigma_r$) from the largest scale $L$ up to the scale $r$ translates into $N > \frac{\Delta v}{\sigma_L}\left( \frac{L}{r} \right)^{m/2}$.
Thus, we do not expect a pure power-law result for a slice of a given width. Indeed,
the largest scales, on the order of the size of the cube, are almost always in a {\it thin} regime; unless, $ N < \frac{\Delta v}{\sigma_L} \sim 5-6$. Figure \ref{fig:ps_rho_11_3_vz_all} is consistent with this behaviour. 

In Figure \ref{fig:ps_rho_11_3_vz_all} we can see that the power spectrum has an additional deviation from the power-law at small scales. For that reason, to measure numerical values of the spectral indices, we have restricted ourselves to the wave-number range from $\log k \sim 1.3$ to $2.2$, where the power spectra in Fig. \ref{fig:ps_rho_11_3_vz_all} is close to a power-law. The slopes (spectral indices) of the power spectra obtained with
a linear least-squares method are shown in Table \ref{tab:ps_all}.
\begin{table}
\caption{Measured emissivity spectral indices of Fig. \ref{fig:ps_rho_11_3_vz_all}}
\label{tab:ps_all}
\begin{tabular}{ccccccc}
\hline
\ & \multicolumn{3}{c}{Density index $n=-2.5$} & \multicolumn{3}{c}{Density index $n=-11/3$}\\
Number &   \multicolumn{6}{c}{Velocity index ($\mu$)}\\
of & -3.3 & -3.67 & -4.0  & -3.3 & -3.67 & -4.0\\
velocity &\multicolumn{6}{c}{Predicted spectral index ($\gamma$)} \\
channels & -2.35 & -2.17 & -2.00 & -2.67 & -2.6& -2.5\\
\hline
150 & -2.02 & -1.97 & -1.93 & -2.25 & -2.19 & -2.15 \\
100 & -2.05 & -2.00 & -1.95 & -2.27 & -2.22 & -2.18 \\
75  & -2.07 & -2.02 & -1.97 & -2.30 & -2.24 & -2.20 \\ 
50  & -2.11 & -2.06 & ~-2.02$^*$ & -2.35 & -2.29 & -2.25 \\
25  & -2.22 & ~-2.18$^*$ & -2.15 & -2.54 & -2.47 & ~-2.44$^*$ \\
15  & ~-2.34$^*$ & -2.33 & -2.33 & ~-2.74$^*$ & ~-2.74$^*$ & -2.74 \\
10  & -2.44 & -2.42 & -2.46 & -3.01 & -2.99 & -3.02 \\
6   & -2.47 & -2.52 & -2.56 & -3.13 & -3.23 & -3.39 \\
4   & -2.47 & -2.53 & -2.55 & -3.13 & -3.29 & -3.44 \\
2   & -2.46 & -2.52 & -2.54 & -3.13 & -3.30 & -3.45 \\
1   & -2.49 & -2.49 & -2.49 & -3.65 & -3.65 & -3.65 \\
\hline
\end{tabular}
\\ $^*$ With this velocity resolution (number of channels) we get the better match for the analytical predictions in thin channels.
\end{table}

From Fig. \ref{fig:ps_rho_11_3_vz_all} (and Table \ref{tab:ps_all}) we can see, indeed, a steepening of the power-spectrum as we increase the thickness of the velocity slices. However, the spectral-indices in thin channels do not appear to reach an asymptotic value, instead they keep getting shallower as the velocity resolution increases. This behaviour arise due to the limited number of emitters along the LOS: with only $216$ emitters along each LOS, to distribute in a large number of velocity channels, there will be many channels empty. This produces artificial  sharp edges, and introduce high frequency components. Rendering in a  shallower spectrum at large wave-numbers. To avoid this problem, in \citet{LPVP} it was suggested as optimal to search for thin slice asymptotics in the channels that are just thin enough,
\begin{equation} 
\label{criteria}
N \approx \frac{\Delta v}{\sigma_L}\left( \frac{L}{r} \right)^{m/2} 
\end{equation}
at the smallest scale $r$ of interest.
The expression in eq. (\ref{criteria}) is somewhat tricky, because we need to know the value of $m$, to get the number of channels to be used to best determine $m$! But, it allows to check with emulated observations if the predictions hold. Our results, in Table \ref{tab:ps_all}, are in agreement with the prediction in \citet{LPVP} that more channels are required when steeper velocity fields are studied. For real observations the problem is alleviated because in that case we are not limited by the number of emitters and we can use the maximum velocity resolution without having empty channels. In \citet{LPVP} the factor of $\Delta v/\sigma_L$ in eq. (\ref{criteria}) was estimated to be $\sim 2.5$, in the assumption of Gaussian statistics. In practice we can use the standard deviation of the velocity in the LOS velocity cube for $\sigma_L$, and $\Delta v$ is the range of velocity covered in our PPV cubes. This is important to keep in mind because a side effect of the modification of the spectral index is an appreciable change in the velocity range of the velocity cubes. To get the factor ($L/r$) we need to take the maximum wave-number used to obtain the power-spectra. For the results presented in Table \ref{tab:ps_all} it was $k \sim 10^{2.2}$, so $L/r=k/2\pi \sim 25$. This way we generated PPV cubes with the velocity resolution according to eq. (\ref{criteria}), and measure the power spectrum in velocity channels, in Table \ref{tab:barely_thin} we display the results.
\begin{table}
\caption{Emissivity spectral indices in barely thin chanels}
\centering
\label{tab:barely_thin}
\begin{tabular}{ccccc}
\hline
Density & Velocity & Num. of & Predicted & Measured \\
index & index & channels & $\gamma$ & $\gamma$ \\
\hline
-2.5  & -3.3  & 35 & -2.35 & -2.2\\
-2.5  & -3.67 & 42 & -2.17 & -2.1\\
-2.5  & -4.0  & 54 & -2.00 & -2.0\\
-3.67 & -3.3  & 35 & -2.85 & -2.4\\
-3.67 & -3.67 & 42 & -2.67 & -2.3\\
-3.67 & -4.0  & 54 & -2.50 & -2.2\\
\hline
\end{tabular}
\end{table}
From Table \ref{tab:barely_thin} we can see that the measured spectral indices within $15\%$ error of the predictions, but always shallower than expected. This is an indication, that the noise arising from the limited number of emitters is present even in channels barely thin, although not near as large as that present for PPV cubes constructed with more channels (see Table \ref{tab:ps_all}).

\begin{figure}
\includegraphics[width=84mm]{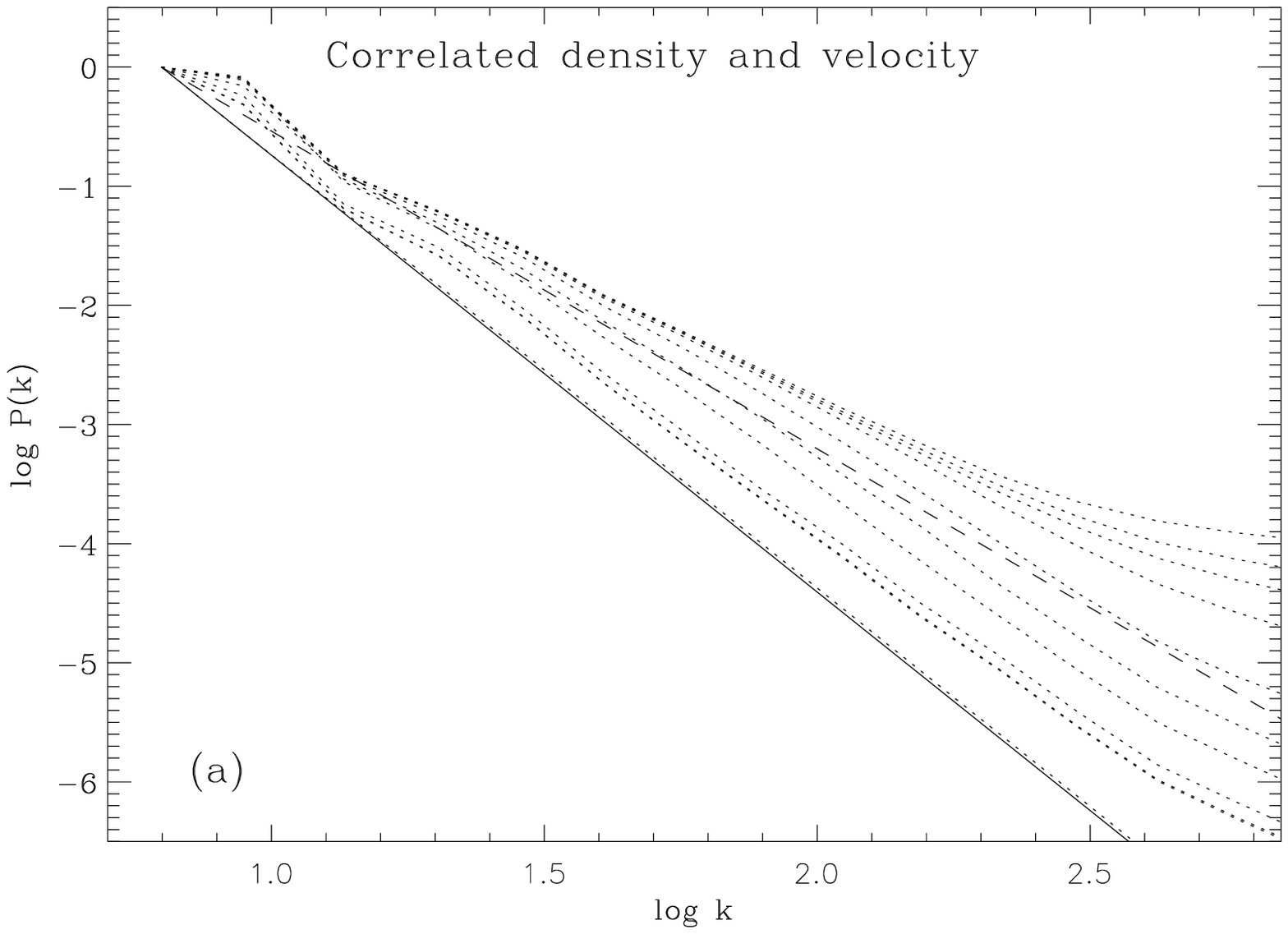}
\includegraphics[width=84mm]{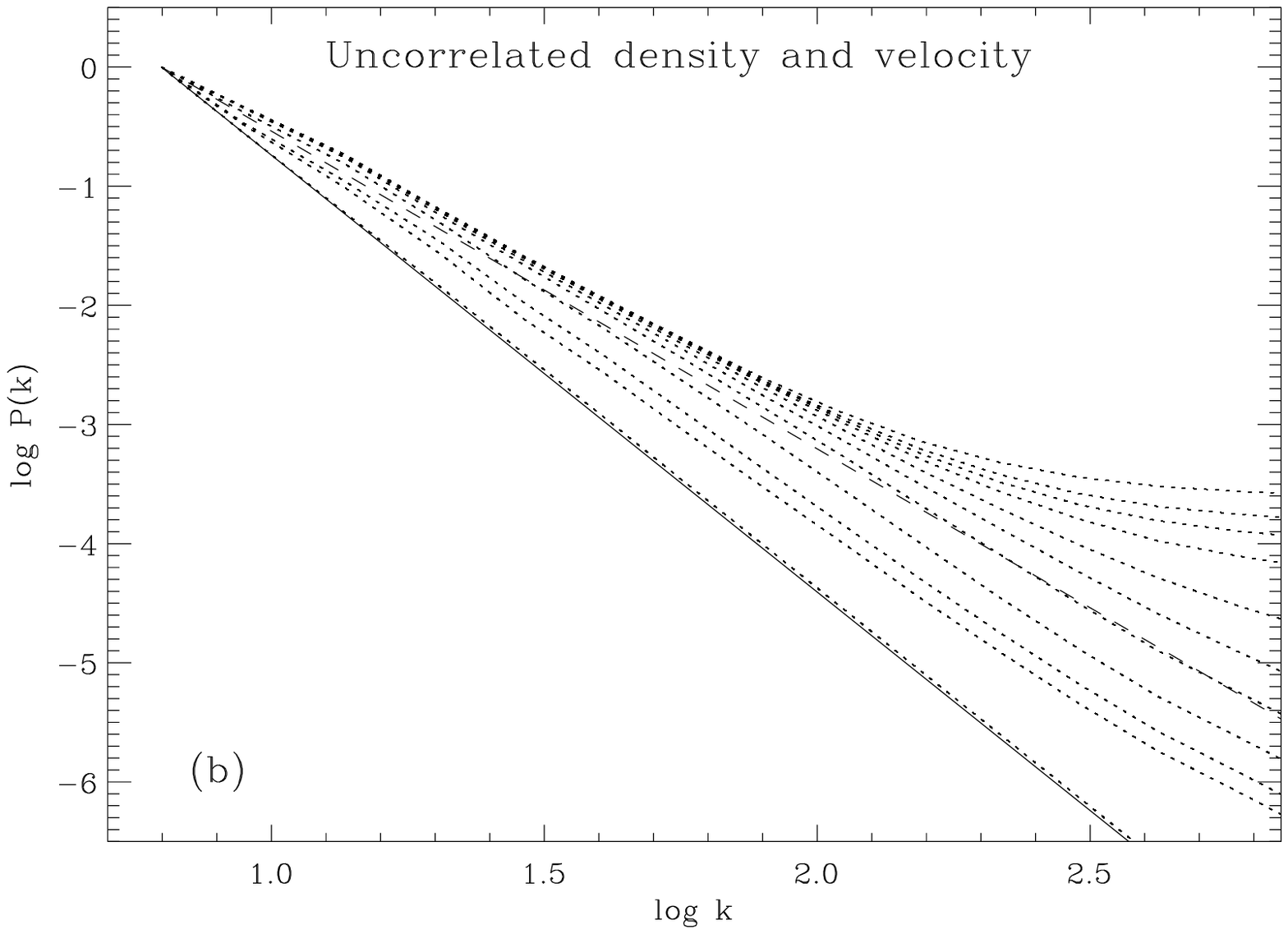}
\caption{Power spectrum in velocity channels for PPV produced with density and velocity cubes from MHD simulations --with spectral indices modified to $-11/3$-- (\emph{a}), and with the same density as the \emph{upper} panel but an ensemble of Gaussian cubes with the same $-11/3$ index (\emph{b}). The \emph{dotted lines} correspond to the average spectrum in velocity channels for PPV with different number of channels, in both panels from these correspond from top to bottom to $150$, $100$, $75$, $50$, $25$, $15$, $10$, $6$, $4$, $2$, and $1$ channels respectively. We see larger departures from power-law behaviour at lower $k$ in ({\it a}) because in ({\it b}) we averaged over an ensemble, see discussion on the text. Also, in both cases for reference we plotted the predictions of the power spectrum in \emph{thick} channels (\emph{solid lines}), and in \emph{thin} channels (\emph{dashed lines}).\label{fig:comp_corrs}}
\end{figure}

Could deviations above be due to correlations between the density and the velocity fields? The answer is no: to illustrate this point, we can compare the power spectrum of the MHD simulations to that of an ensemble of synthetic PPV cubes with the same spectral indices but no velocity-density correlations whatsoever. 
For this comparison to work we must use PPV cubes constructed from data with the same power-law inertial range. For that reason, we modified the spectral index of the density field from the simulations to $-11/3$ (close to the original index, so the modification on the velocity-density correlations is minimal). With this density field, we then constructed PPV cubes using the modified velocity output from the simulations and $60$ synthetic velocity fields. All the velocity cubes used had the same spectral index (also $-11/3$). The power spectra in velocity channels is shown in Fig. \ref{fig:comp_corrs}.
In this figure we can appreciate that apart of some `wiggles' at small values of $k$, due to lack of enough statistics (see discussion below) the power-spectra is practically identical whether the density field is correlated with the velocity field or not.

\subsection{More on the numerical limitations}

As we saw in the previous section, one important practical problem with the determination of the spectral indices through VCA is to measure with accuracy the slopes (precisely the spectral indices) in the plots. We can notice from each plot in Fig. \ref{fig:ps_rho_11_3_vz_all}, two departures from well behaved power-laws. At the left end of the spectrum (low $k$ numbers, large spatial scales) we have some random wiggling, while at the right end of the curves we see a systematic flattening of the power spectrum. The latest is increasingly evident for PPV cubes formed with larger number of channels (with thinner velocity channels), and was referred as `shot-noise' in \citet{LPVP}.

\subsubsection{Departures from power-law at low wave-numbers}

\begin{figure*}
\includegraphics{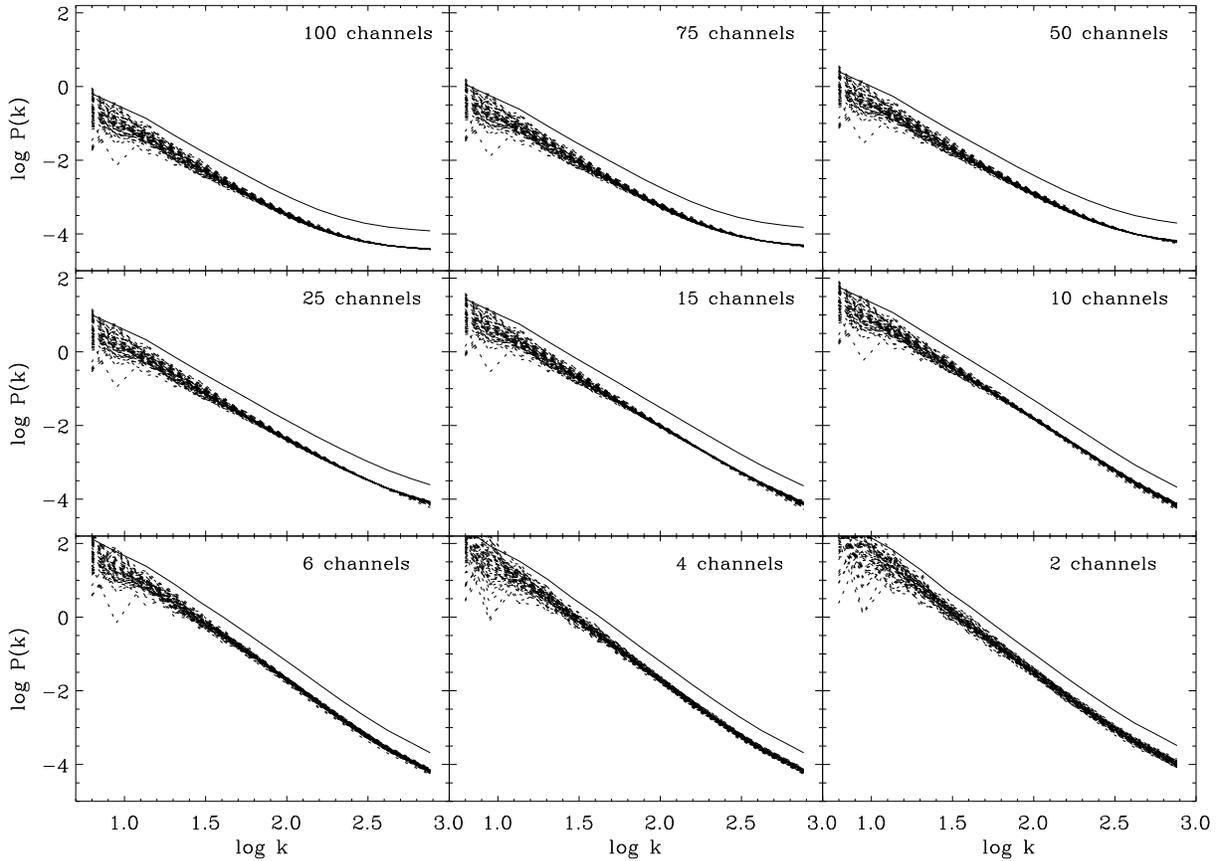}
\caption{Power spectra in velocity channels for an ensemble of $60$ Gaussian cubes of velocity with index $-11/3$ and constant density. The \emph{dotted lines} are the average of the 2D power spectra in velocity channels for individual PPV cubes constructed with the number of channels as denoted in the upper right corner in each plot The \emph{solid lines} correspond to the ensemble average, for visual purposes they are shifted vertically by $0.5$.}
\label{fig:ens_singles}
\end{figure*}

In the spirit of avoiding other sources of uncertainty, we will use simple Gaussian fields with prescribed power-law spectrum as velocity fields, mapped with a constant density field to produce PPV cubes. We generated an ensemble of $60$ Gaussian cubes with a prescribed spectral index of $-11/3$.  Then we applied to them the VCA, that is, we obtained PPV cubes with different number of channels (different velocity resolution) and compute the 2D power spectra in velocity channels. The results are shown in Fig. \ref{fig:ens_singles}.
From the figure is evident that these departures from strict power-laws at low $k$ numbers are due to the lack of sufficient statistics at the largest scales. Power spectra in velocity channels for individual velocity cubes differ very little from intermediate to large $k$ numbers, but differ significantly at small $k$, showing some `wiggles'. However, the ensemble average converges to a power-law that extends from the largest spatial scales to intermediate spatial scales, until a systematic flattening of the power spectrum at large $k$ becomes evident. This problem may ocurr in actual observations, where the low wave-numbers (large scales) are affected by the general size and shape of the object under study.

\subsubsection{Departures from power-law at high wave-numbers}

One of the most important limitations that we to construct spectral lines with numerics, is the finite (and very small) number of emitters in a particular direction; in our simulations we have $216$ density elements in each LOS to distribute in velocity channels. If we impose a high velocity resolution in the PPV cubes (i. e. large number of channels), the spectral lines instead of being smooth will have sharp edges and empty channels. This artificial jumps introduce high frequency components in the spectrum and therefore an increase of the power spectrum at large wave-numbers. In \citet{LPVP} this was regarded as `shot-noise'. This is true in the sense that the flattening becomes larger as we increase the velocity resolution. But is more complicated that the usual shot-noise, that can be easily described in terms of Poisson statistics. This flattening of the spectrum is particularly problematic because it increases as we go into thinner channels, and the predictions in LP00 to extract the velocity spectral index were done precisely in an asymptotic regime for thin channels.

\begin{figure}
\includegraphics[width=84mm]{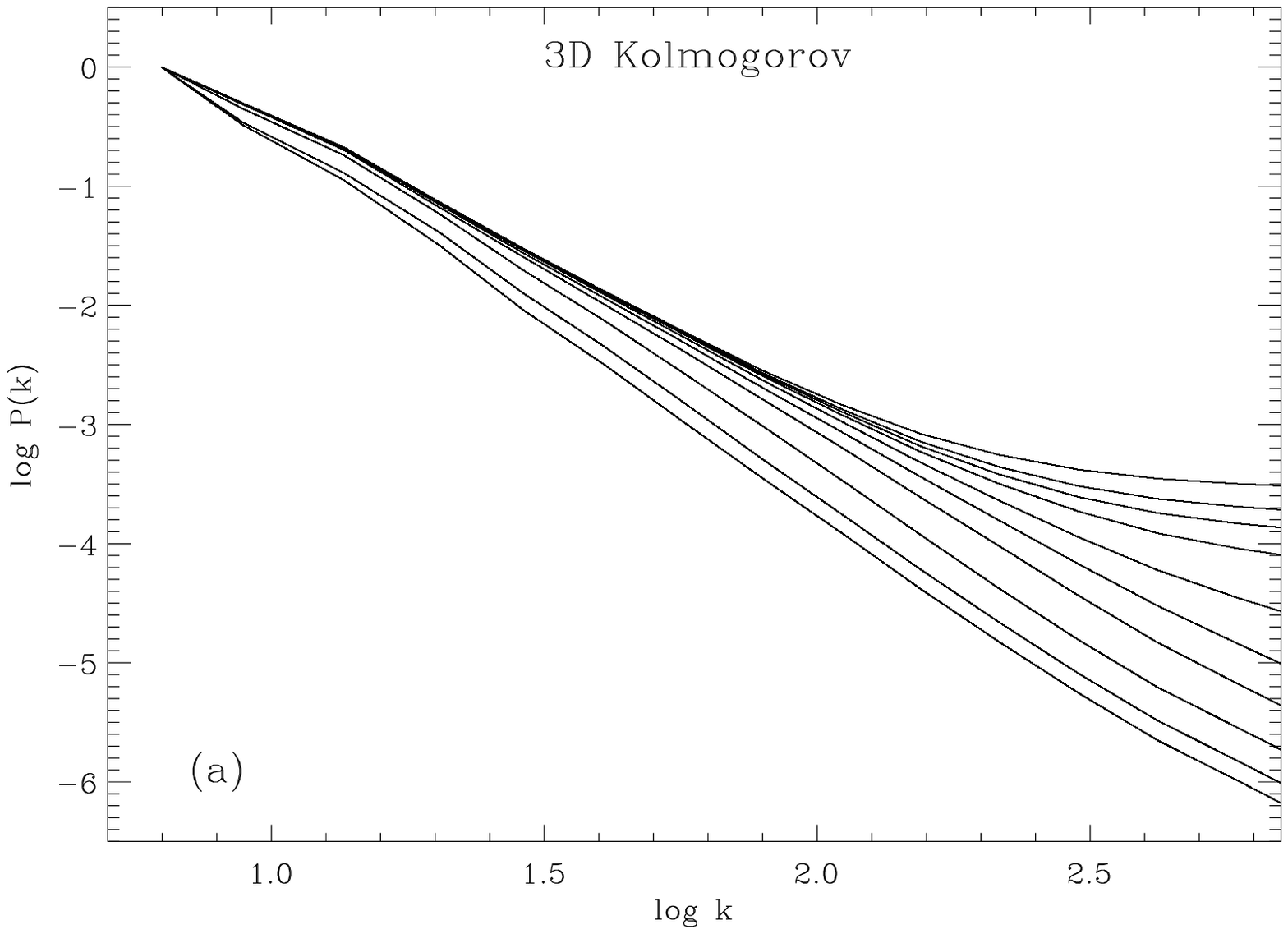}
\includegraphics[width=84mm]{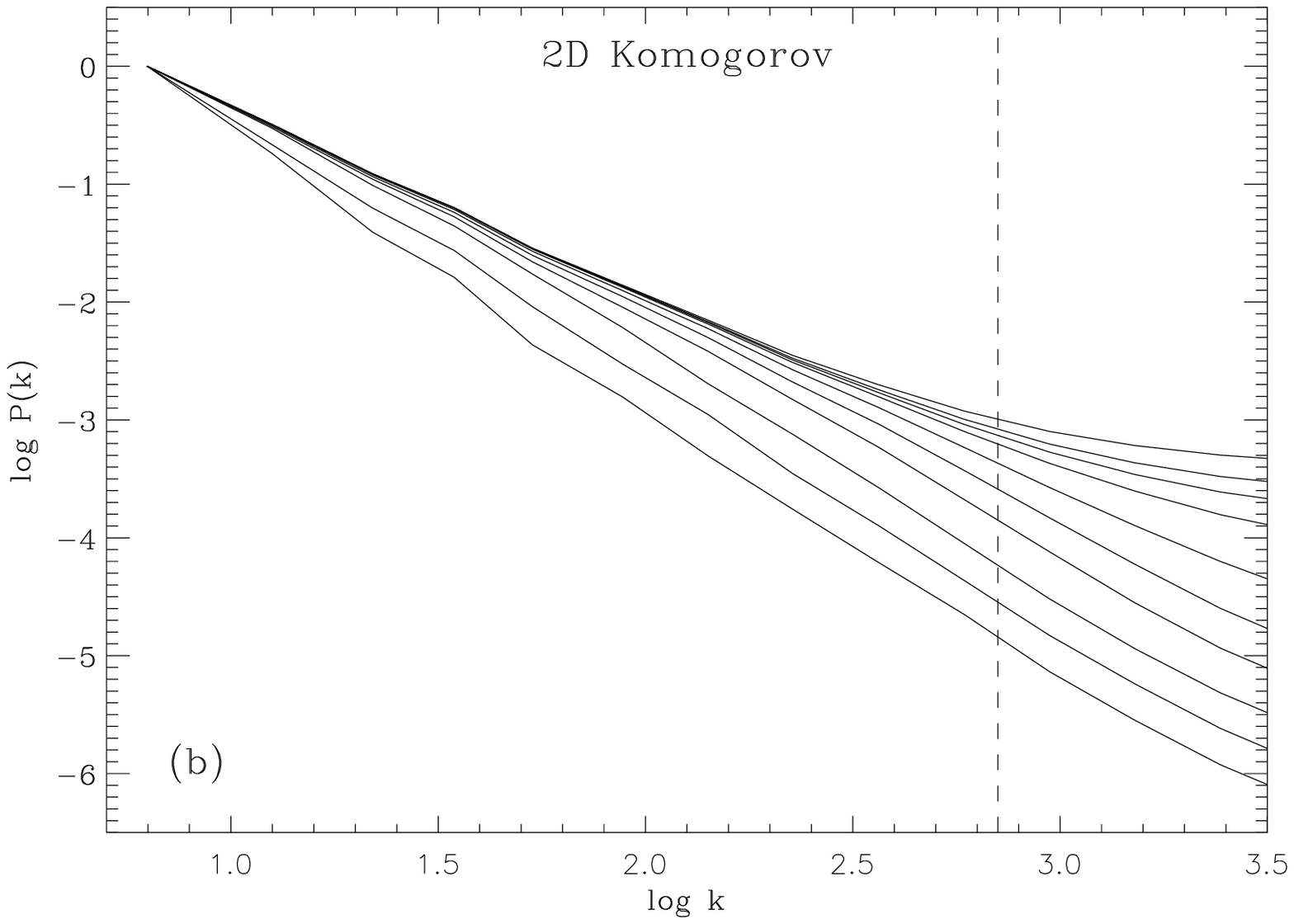}
\caption{Comparison of the power spectrum in velocity channels for an ensemble of $60$ velocity cubes ({\it a}) with an ensemble of $50$ squares ({\it b}). The cubes are $216^3$ in size and the squares are $1536^2$. All velocity fields have Kolmogorov spectral indices ($-8/3$ for 2D and $-11/3$ for 3D), and the PPV were obtained with a constant density. The dashed line on the lower panel is to compare the inertial range achievable in both cases.}
\label{fig:ens_2d_3_d}
\end{figure}

A natural way to avoid the problem of this flatening of the power spectrum is simply to increase the number of emitters along the LOS. This can be accomplished increasing the size of the data cubes, which translates into more computing power required to analyse the power-spectrum (and more importantly more computing power to generate larger MHD cubes). 
With the resources at our hand, we increased the number of emitters along each LOS by creating an ensemble of $50$ 2D Gaussian velocity fields of size $1536^{2}$. All with spectral index of $-8/3$, Kolmogorov in 2D. Analogously to the 3D ensemble we simulated observations and analysed the power spectrum in velocity slices. In this case instead of PPV cubes we have simply `PV arrays', and the power spectrum computed in slices is one-dimensional. In Fig. \ref{fig:ens_2d_3_d} we compare the ensemble average power-spectrum in velocity channels for both, 3D and 2D cases.
It's clear the increase in wave-number range acquired with the 2D data, in which we have about half decade more. The relevant result of this comparison is to show that the flattening of the power spectrum for the thinnest channels is present in both cases, at large $k$. But for the 3D case this flattening is present earlier than in the 2D case.

It's important to stress that this is NOT a real dynamical effect, the cubes used to construct the PPV cubes are just Gaussian velocity and constant density.

\begin{figure}
\includegraphics[width=84mm]{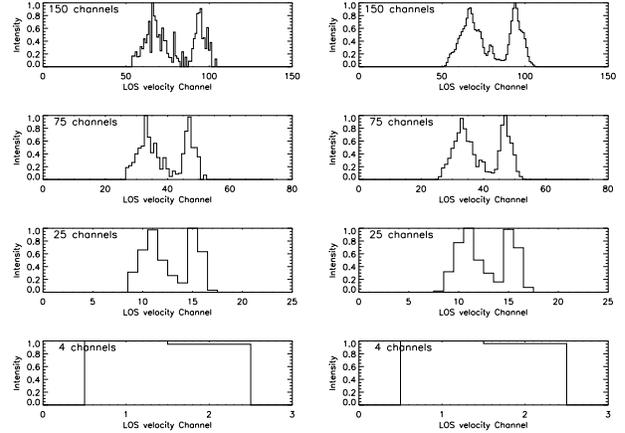}
\caption{Examples of spectral lines in one of our PPV cubes with different resolutions (the number of channels is indicated in each panel.). On the {\it left} we have spectra from PPV cubes with no width associated to the emitting elements, on the {\it right} the same spectra after adding a significant width (Gaussian with FWHM of $\Delta /60$). All the velocity axis contain the entire velocity range in the PPV, and the intensities are normalized for visual purposes.}
\label{fig:s_lines}
\end{figure}

Another way to alleviate the problem of limited number of emitters is to include an intrinsic width to each emitting element. We acomplished this task by adding some Gaussian smoothing on each spectral line (i. e. convolving a Gaussian profile along each LOS). This is somewhat equivalent to introduce a thermal width to the emitters\footnote{We used this width as free parameter, therefore we rather use the term `intrinsic' instead of `thermal' width, wich stricltly speaking, is determined by physical mechanisms beyond the scope of our simulations.}. Some examples of the spectral lines in our PPV cubes, with and without an intrinsic width are shown in Figure \ref{fig:s_lines}.
Introducing an intrinsic width to the PPV cubes makes the spectral lines to look more realistic because it smooths sharp edges due to the limited numerical resolution. However, it also limits the resolution that can be used to look for thin channel asymptotics. The criterion in LP00 for {\it thin} or {\it thick} channels, was based on an `effective width' of the velocity channels, which included not only the thickness of the channel but also a thermal width $v_T$ as:
\begin{equation}
\delta v_{eff} \approx \left( \delta v^2+2v_T^2 \right)^{1/2}.
\end{equation}

\begin{figure}
\includegraphics[width=84mm,height=84mm]{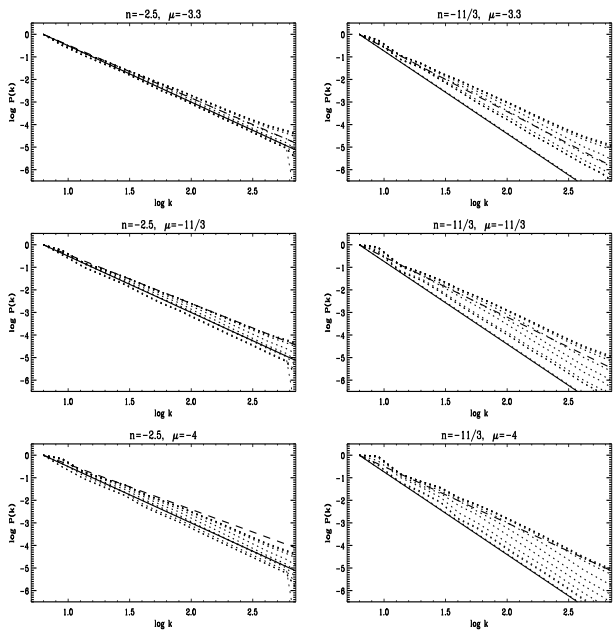}
\caption{Same as Fig. 2 but including an intrinsic width to the emitters. Each emitting element is a Gaussian with a FWHM of $\Delta v/60$. We can see how the flatening of the power spectrum at large $k$ decreases dramatically.}
\label{fig:ps_smooth}
\end{figure}

The additional intrinsic width limits the resolution attainable in two ways. For a fixed maximum $k$ scale the velocity channels need to be narrower to remain in the {\it thin} regime, with a minimum number of channels given by:
\begin{equation}
\label{eq:criteria_vt}
N \gtrsim \left[ \frac{\sigma_L^2}{\Delta v^2}\left(\frac{r}{L}\right)^m-2 \frac{v_T^2}{\Delta v^2}\right]^{-1/2}.
\end{equation}
And for a fixed velocity resolution it reduces the wave-number up to which the channel remains {\it thin} to
\begin{equation}
\label{eq:criteria_k}
k \lesssim 2 \pi \left[ \frac{1}{\sigma_L^2}\left(\frac{\Delta v^2}{N^2}+2v_T^2\right)\right]^{-1/m}.
\end{equation}
The criteria in eq.(\ref{eq:criteria_k}) is more restrictive, beyond that scale the velocity channels are entering in a transition to the {\it thick} regime. The condition in eq.(\ref{eq:criteria_vt}) is less restrictive, in the sense that any slice with that minimum number of channels is to be considered {\it thin}. However, there is no reason to slice narrower than that, beacuse the thermal width smears velocity fluctuations on smaller scales. If we go to yet thinner channels we reduce the amplitude of the signal, but do not see any further change of spectral index. In Figure \ref{fig:ps_smooth} we show the power spectra in velocity channels for PPV cubes where we included an intrinsic width similar to the width of the barely thin channels (thin down to the scale corresponding to the limit we chose previously to measure the power spectra, $\log k \sim  2.2$).
From the figure we can notice immediately that the rising tail at high wave-numbers dramatically decreases. In the cases where the velocity fluctuations at small scale structure are relativelly small (in Fig. \ref{fig:ps_smooth}, $\mu =-4$) the transition from thin to thick channels as we go into smaller scales (large $k$) is evident. This corresponds to predictions in LP00.
We computed the spectral indices in velocity channels, for our previous inertial range with slices of thickness according to eq.(\ref{eq:criteria_vt}), and also for the conservative inertial range from eq.(\ref{eq:criteria_k}) for the same velocity resolutions as before. In both cases we observe that the measured spectral indices doesn't change significantly from the earlier estimates with barely thin channels (presented in Table \ref{tab:barely_thin}). This is understandable because the intrinsic width we included alleviates the problem of small number of emitters only on scales smaller than such width.

This shot-like noise presents an obstacle if we want to obtain the velocity index of data cubes generated synthetically. However, we must stress that the problem is not expected to be present in real observations, where the number of emitters along the LOS is much larger. This was confirmed through observations of the Small Magellanic Cloud \citep{SL}.  The spectral indices there, measured in thinner and thinner channels do tend to an asymptotic value, not being affected by `shot-like noise' allowed to measure accurately the power-spectrum in velocity channels.

\section{Effect of the shear on the VCA}

\begin{figure}
\includegraphics[width=84mm]{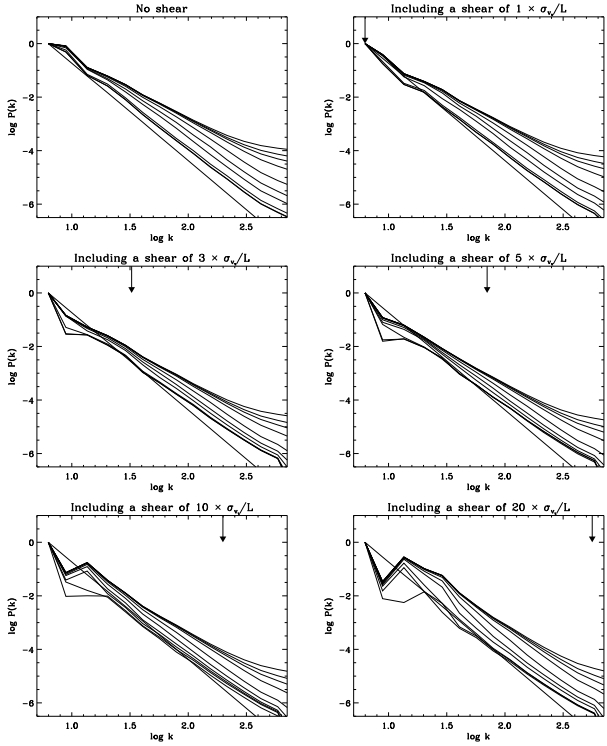}
\caption{Power spectra in velocity channels including linear shear. Each panel correspond to a different magnitude of the maximum shear, as labelled in the title of each plot. The different lines in each panel correspond from top to bottom to the average power spectrum in velocity channels for PPV cubes constructed with $150$, $100$, $75$, $50$, $25$, $15$, $10$, $6$, $4$, $2$, and $1$ channels respectively. The small arrows show at which wave-number the magnitude of the external shear equals the magnitude of the turbulent shear.\label{fig:shear}} \end{figure}

The VCA is sensitive to density and velocity fluctuations at a particular scale. On the other hand, shear as any organized motion, like galactic rotation, would correspond to the smaller wave-numbers (the largest scales, sometimes much larger that the largest scale in a PPV cube). If we choose the right range in $k$ to perform the VCA, we should avoid any interference of this ordered motions with the analysis. To explore numerically the effect of a linear shear on the VCA we used modified data cubes of spectral indices of $-11/3$ for both velocity and density fields. We chose the LOS to be along the $z$ axis and added a component to the velocity cube (the $v_{z}$ cube) following
\begin{equation}
\label{shear}
v_{z}(x,y,z)=v_{0z}(x,y,z)+C\sigma_{v_{z}}\frac{x}{L}\hat{\bmath{z}},
\end{equation}
where $v_{z}$ is the new velocity cube used to generate the PPV cubes, $v_{0z}$ is the original velocity cube before introducing the shear, $\sigma _{v_{z}}$ is the velocity dispersion of $v_{0z}$, $L$ is the cube size in the spatial coordinates ($x$, $y$), and $C$ is a parameter we introduce to vary the magnitude of the shear introduced. The shear introduced is the same on all scales, and equal in magnitude, to $C \sigma_{v_z}/L$. We generated PPV cubes adding a linear shear with magnitude up to $1$, $3$, $5$, $10$, and $20$ times maximum velocity dispersion over the largest scale (i. e. with a velocity component with $C=1$, $3$, $5$, $10$, and $20$ in eq. \ref{shear}). As before, we measure the spectral indices in the range less affected by numerical noise, $\log k$ from $\sim1.3$ to $\sim2.2$. The range used is somewhat arbitrary, but the purpose of the study at this point is just comparison with the case of no shear. The measured slopes are summarized in Table \ref{tab:shear}.
\begin{table}
\caption{Spectral indices in velocity channels with shear}
\label{tab:shear}
\begin{tabular}{lcccccc}
\hline
Num. of~  & No & \multicolumn{5}{c}{Shear magnitude (in units of $\sigma_{v_z}/L$)}\\
channels& shear& $1$& $3$& $5$&$10$&$20$ \\
\hline
150& -2.2& -2.3& -2.4& -2.4& -2.7& -3.1 \\
100 & -2.3& -2.3& -2.4& -2.5& -2.7& -3.1 \\
75 & -2.3& -2.4& -2.5& -2.5& -2.8& -3.2 \\
50 & -2.4& -2.4& -2.6& -2.6& -2.9& -3.3 \\
25 & -2.6& -2.7& -2.8& -2.9& -3.2& -3.7 \\
15 & -2.9& -3.0& -3.0& -3.1& -3.4& -4.1 \\
10 & -3.2& -3.1& -3.1& -3.3& -3.4& -4.0 \\
6 & -3.4& -3.3& -3.1& -3.1& -3.3& -3.7 \\
4 & -3.4& -3.3& -3.1& -3.1& -3.3& -3.5 \\
2 & -3.4& -3.3& -3.1& -3.1& -3.2& -3.4 \\
1 & -3.7& -3.7& -3.7& -3.7& -3.7& -3.7 \\
\hline
\end{tabular}
\end{table}
The introduced shear is the same on all scales, while the turbulent shear scales as $\sim k^{2/(m-2)}$. The magnitude of the external shear is the same as the magnitude of the turbulent shear when $k \approx 2\pi C^{3/2}$. Therefore the external shear contributed initially to the power spectrum only at the largest scales, but eventually became important at intermediate to small scales as its magnitude increased (see Fig. \ref{fig:shear}).
Our results show that the effect of an external shear becomes important only when it is comparable to the local turbulent shear. In our example that was when the maximum shear reached about $5$ times the velocity dispersion over the largest scale. The relative differences of the spectral indices measured compared with the case of no shear are below $10\%$, for the cases of less shear ($C \lesssim 5$). It is noticeable that the larger distortion happens at lower $k$, even when its magnitude is comparable to the turbulent shear at smaller scales. This is expected, given that the shear is regular, and represents a large scale motion. And holds if we measure the slope of the power spectrum in intermediate spatial scales, which are available using synthetic data. With actual observations, we can go further into smaller scales and avoid the effects of a stronger shear. The shear due to turbulence dominates in the galaxy, and its the effect to the VCA is marginal.

\section{Anisotropic turbulent cascade}

Magnetic field plays a crucial role in the dynamics of interstellar turbulence, and makes the turbulent cascade anisotropic \citep{M82, H84}. In a turbulent magnetized media, the energy on the large scale motions is larger than that on the small scale structures; however the local magnetic field strength remains almost the same. Thus, is easier to bent magnetic field lines on the large scale by turbulent motions, but on the small scale becomes more difficult.
Another way to think about this phenomenon is noting that hydrodynamic motions can easily mix magnetic field lines on the small scale, rather than bent them, in the direction perpendicular to the mean magnetic field (see a discussion in \citealt{LV}). As a result we get elongated eddies, relative to the magnetic field that become even more elongated as we go further into smaller scales. A model for incompressible anisotropic turbulence is that of \citet{GS95}. This model has been supported by numerical simulations (e.g. \citealt*{CV00, MG, CLV}). Compressible anisotropic MHD turbulence has been studied in \citet{LG} and \citet{CL,CL03}.

\subsection{Magnetic Field Direction: Anisotropy in Correlation Functions}

In an isotropic turbulent cascade correlations depend only on the separation between points. Contours of equal correlation are circular in that case. If on the other hand the turbulence is anisotropic, two point statistics are also anisotropic. In this case contours of iso-correlation become elongated, with symmetry axis given by the magnetic field direction. The technique presented here is similar to one proposed to study magnetic field direction using synchrotron maps \citep{L92}. Iso-contours of correlation functions can allow us to determine the direction of the magnetic field. We calculate two point correlation for different 2D maps generated with our data. The correlation function between the two scalar functions $f(\bmath{x})$ and $g(\bmath{x})$ can be computed from
\begin{equation}
\label{crcorr}
C(\bmath{r})=\frac{\langle (f(\bmath{x})-\langle f(\bmath{x})\rangle )\bmath{\cdot} (g(\bmath{x}+\bmath{r})-\langle g(\bmath{x})\rangle )\rangle } {\sigma _{f}\sigma _{g}},
\end{equation}
where $\bmath{r}$ is the separation between two points or `lag', $\bmath{x}$ is the spatial separation, $\sigma _{f}$ and $\sigma _{g}$ are the standard deviations of $f$ and $g$ respectively. Here the average is performed on all the radial components and azimuthal angles.

\begin{figure}
\includegraphics[width=84mm]{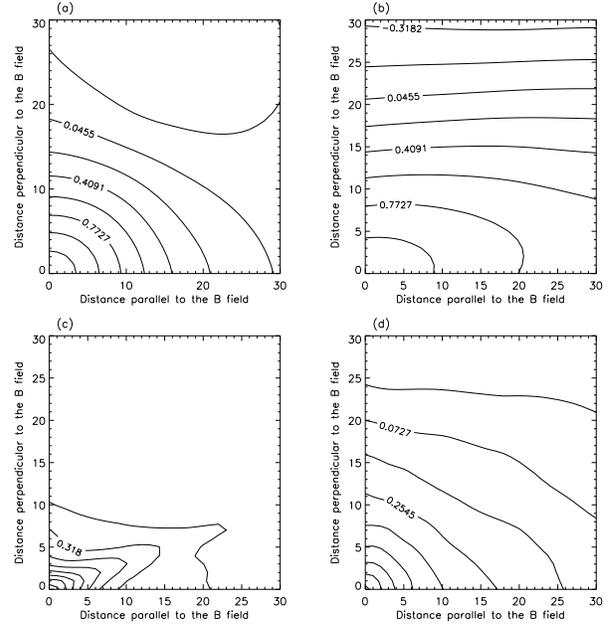}
\caption{Contours equal correlation of 2D maps obtained with the PPV data. ({\it a}): contour map of the correlation of emissivity at the centroid of the lines. ({\it b}): contours of correlation in the integrated emissivity. ({\it c}), ({\it d}): contours in two individual channel maps. These were obtained in a PPV cube with 25 channels (the velocity range is from $V_{min}=-1.27$ to $V_{max}=1.451$, and consequently the channel thickness is $\delta V=0.109$).
({\it c}) corresponds to a velocity centred at $V_{z}=-1.11\protect$, while the ({\it d}) to a velocity of $V_{z}=0.741$. All the distances in the figures are in grid units.\label{fig:corrs}}
\end{figure}

With the original cubes from simulations (no artificial modification of the spectral indices), we calculated the auto-correlation (correlation in eq. (\ref{crcorr}) with $f(\bmath{X})=g(\bmath{X})$) on various 2D maps: integrated intensity (integrated along the LOS, equivalent to a 2D map of column density); emissivity at the centroid of the lines, obtained as:
\begin{equation}
\label{centroid}
j_{c}(\bmath{X})=\frac{\int j(\bmath{X})v{\rm d} v}{\int j(\bmath{X}){\rm d} v},
\end{equation}
where the emissivity $j(\bmath{X})$ in our case is proportional to the first power of the density; and maps of emissivity in individual velocity channels.
Contour maps of iso-correlations calculated of these maps are given in Fig. \ref{fig:corrs}.
We can see that anisotropy is present in the contours plotted. It is evident in the case of the emissivity at the centroids of the lines, as well as in the integrated emissivity. In individual velocity slices, we found that the majority of the velocity channels show this anisotropy in the expected direction (elongation of the contours parallel to the mean B field), but in some slices we see the elongation of correlation contours tilted from the mean magnetic field direction. For Fig. \ref{fig:corrs} we choose a representative slice to illustrate this misleading behaviour.

The main purpose of this subsection was to show an alternative method to obtain the direction of the magnetic field, and also to illustrate that anisotropic cascade is present in our simulations.

The anisotropy based technique could be a unique tool for magnetic field studies, especially when other techniques, e.g. based on dust alignment fail (see \citealt{L00} for a review on grain alignment).

\subsection{Anisotropic cascade \& VCA.}

\begin{figure*}
\includegraphics{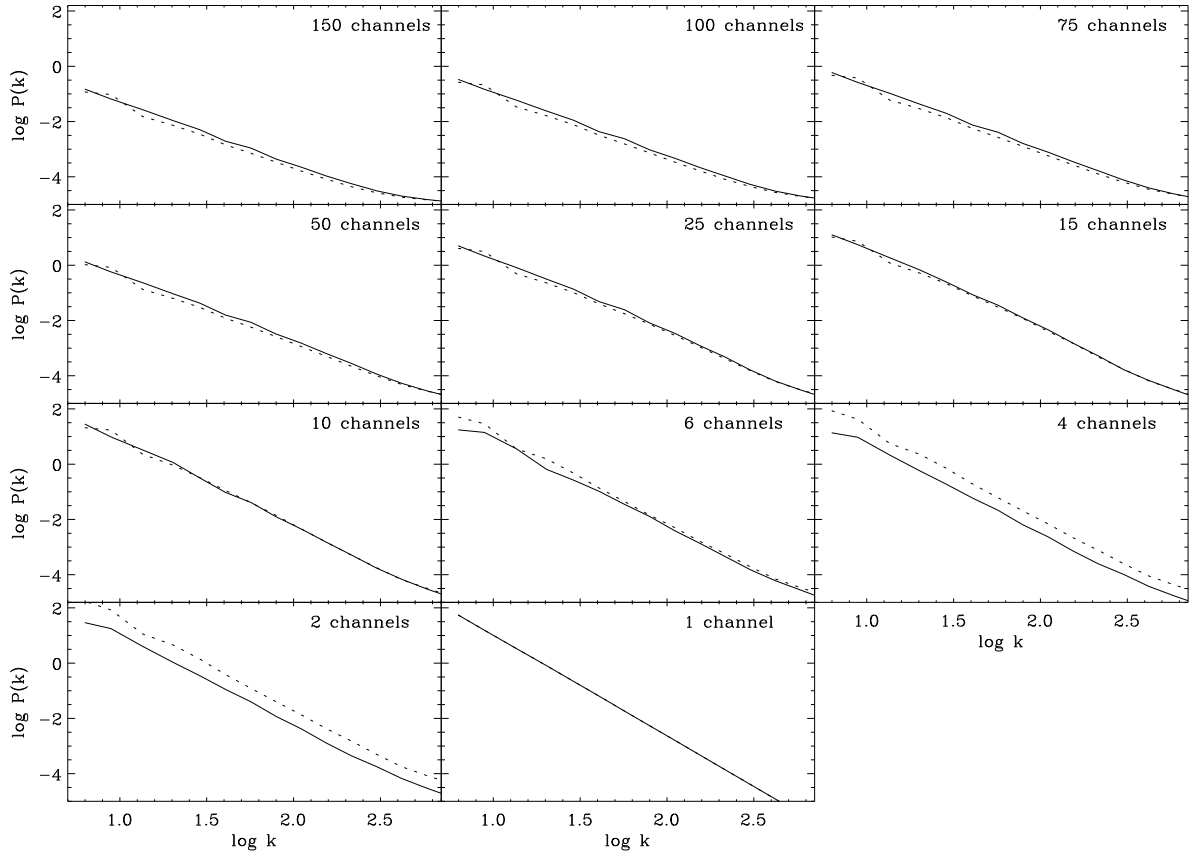}
\caption{Power spectra in velocity channels for observations with the observer at two different positions. Each panel correspond to the average spectrum in PPV cubes with different number of velocity channels (the number of channels is indicated in the upper right corner on each). The {\it solid lines} correspond to the LOS parallel to the mean magnetic field; for the {\it dotted lines} the LOS is perpendicular to the mean magnetic field. All the PPV cubes were constructed from the same set of density and velocity data, `modified' to 3D spectral indices of $-11/3$.
\label{fig:ps_vz_vx}}
\end{figure*}

It's important to test the VCA, which was developed considering isotropic turbulence, in a more realistic case where the magnetic field determines a preferential direction of the turbulent motions. With that purpose we constructed PPV cubes choosing the LOS to be parallel to the mean magnetic field (along the $x$ axis) and compare the results with the case where the magnetic field is perpendicular to the LOS ($z$ axis).
For this exercise we use a data cubes corrected to Kolmogorov spectral indices ($-11/3$ in 3D). 
 The power spectrum in velocity channels is shown in Fig. \ref{fig:ps_vz_vx}.
In the figure we can see that the slopes of the spectrum (spectral index) are almost identical whether the observer is looking parallel or perpendicular to the mean magnetic field. We performed a least-squares linear fitting and found the differences of the slopes in every case are well within the fitting uncertainties.

We conclude that VCA is applicable to anisotropic turbulence studies.

\section{Summary and conclusions}

In this paper we have tested the VCA technique numerically. We used compressible MHD simulations, and an ensemble of Gaussian fields to emulate spectroscopic observations (PPV data cubes). We computed the power spectra in velocity channels to test the analytical predictions in LP00.

To obtain the PPV data we modified the output of the simulations to force power-law indices to obtain density fields with spectral indices of $-2.5$ and $-11/3$, and velocity fields with spectral indices of $-3.3$ to $-4.0$). The spectral indices derived from our emulated observations are within a $15\%$ error of the analytical predictions. We confirmed trends predicted in LP00, like the steepening of the emissivity index as we increase the width of the velocity channels, or that the velocity fluctuations dominate the emissivity for thin velocity channels. This last result is in agreement with a previous study by \citet{PVGPB}, and is important because warns observers against interpret any structure in PPV as density enhancements or `clouds' (see also discussion in LP00).

We showed that the deviations from power-law behaviour at the largest spatial scales (small $k$ numbers) are due to lack of good statistics at those scales. This was comparing the results obtained with PPV produced with a constant density and the modified velocity field of spectral index $-11/3$, to those using an ensemble of $60$ Gaussian cubes with the same index.

An important source of error addressed is a flattening of the power spectrum at large wave-numbers. This is due to the limited number of emitters along the LOS. We illustrated this point comparing the results obtained with the ensemble of $60$ cubes with $216$ emitters along the LOS, and an ensemble (two-deimensional) with $1536$ emitters. We also try including an intrinsic width to the emitters along the LOS to smooth the sharp edges arising from the limited number of emitters. This procedure reduced the flattening of the spectrum at large $k$ numbers. But, also increases the effective width of the channels, reducing our ability to make use of the scales on wich we reduced such flattening of the power spectrum.
This problem was found to be the most restrictive to our work. However, is exclusive of numerical simulations, and is not expected to be important in real observations (like those presented in \citealt{SL}), where the number of emitting elements is essentially infinite.

We introduced linear shear to our data, and showed that its effect only becomes important when the magnitude of the shear is comparable to the local turbulent shear. We showed, that the effect of large scale shear is marginal at small scales. And that avoiding the largest scales to computed the power-spectrum allowed to recover the turbulence statistics. This shows that for ordinary observational situations we can safely neglect shear in the VCA.

We analysed the cross-correlation for 2D maps of integrated intensity and emissivity at the centroids of the lines, and showed that they are anisotropic (elongated), with symmetry axis defined in the direction of the mean magnetic field. This constitutes a new technique to study the direction of the magnetic field. We hope that the anisotropies will reveal the magnetic field within dark clouds, where grain alignment and therefore polarimetry fails. From the point of view of VCA, we illustrated that the effect of the anisotropy of the turbulence is marginal.

\section*{Acknowledgments}
We thank the referee for valuable suggestions that improved this work. AE acknowledges financial support from CONACyT (Mexico). Research by AL and JC is supported by NSF grant AST 01-25544.

\end{document}